\documentclass[cite]{epl}
\usepackage{graphics}
\title{Low-Energy Properties of  a One-dimensional System
of  Interacting Bosons with Boundaries}
\shorttitle{Low-Energy Properties of  $1{\rm d}$ interacting bosons...}
\author{M. A. Cazalilla}
\institute{The Abdus Salam ICTP,
Strada Costiera 11, 34014 Trieste, Italy, and\\
Donostia International Physics Center (DIPC),
Manuel de Lardizabal 4,\\ 
20018 Donostia, Spain.}
\begin{document}
\maketitle
\begin{abstract}
The ground state properties and low-lying 
excitations of a (quasi) one-dimensional system of 
longitudinally confined interacting 
bosons are studied. This is achieved by
extending Haldane's harmonic-fluid  description to open 
boundary conditions.  The boson density,  one-particle density
matrix,  and momentum distribution are obtained accounting 
for finite-size and boundary effects. 
Friedel oscillations are found in the density. Finite-size scaling of
the momentum distribution at zero momentum is  proposed as 
a method to obtain from the experiment the exponent that governs 
phase correlations.  The strong correlations between  bosons 
induced by reduced dimensionality and interactions are displayed by
a Bijl-Jastrow wave function for the ground state,
which is also derived. 
\end{abstract}

The names of Tomonaga and Luttinger are commonly associated 
with interacting electron systems  in one-dimension (1d). The 
so-called Tomonaga-Luttinger
liquids (TLL's)  have the remarkable property that their low-energy
spectrum is completely exhausted by gapless collective excitations, i.e. 
there are no quasiparticles resembling the constituent electrons.  
However, as emphasized  by Haldane~\cite{Haldane81a,Haldane81b}, the class 
of TLL's is  broader than the electron systems
that have attracted much attention, especially in recent years. 
In this paper we shall be concerned  
with 1d systems whose constituent particles are not 
electrons, but bosonic atoms. 
A few experimental systems which fit well  these 
requirements are already  available: Atomic vapors 
confined in highly anisotropic
traps~\cite{Gorlitz01} and the axial phase of $^4$He absorbed 
in narrow pores~\cite{Wada01} or
nanotubes~\cite{Calbi01} are good examples.
The possibility of TLL behavior in trapped vapors
has already been discussed in the
literature~\cite{Monien98,Yip01}.
Monien {\it et al.}~\cite{Monien98} have pointed
out that the ground state correlations 
would decay as power laws, while Yip~\cite{Yip01} 
has found that the absorption line-shape
would  exhibit power-law singularities.   

Motivated by recent experiments~\cite{Gorlitz01,Wada01},
we have considered the low-energy excitations and
ground state properties of {\it finite} quasi-1d interacting boson systems.
It is  important to realize that so far  the experimental samples 
are mesoscopic in size~\cite{boundaries} and
longitudinally confined. The experimental consequences 
of this fact for interacting 1d {\it bosons} have not been 
fully addressed before.
Nevertheless, it is known~\cite{Fabrizio95,Eggert96}
that the low-energy spectrum and correlation
functions of a finite {\it fermionic} TLL with  open boundaries
are very different. In contrast to a system well
described with periodic boundary conditions, such as 
a quantum degenerate gas in a tight toroidal trap (a ``boson ring"),
a bounded fluid  cannot sustain quantized
persistent  flows, as a ring would, and 
its excitations  are standing  waves.
Even more important, 
Eggert {\it et al.}~\cite{Eggert96} have shown that the 
exponents governing correlations in 
a finite bounded  fermionic TLL cross over very slowly
to the exponents of the infinite system. 
The differences  illustrate the effect 
of boundaries on the quantum critical fluctuations 
characteristic of these systems at very low temperatures,
and as we shall argue, they should be
taken into account when confronting
the experiment.  Before we proceed any further,
it should be noted that in this letter the longitudinal confining
potential is approximated by two infinite barriers at the 
boundaries. This is not generally the case in current experiments with
trapped atomic vapors, where the confining potential is usually 
harmonic. Accounting for this fact within
the formalism employed in this paper  is not a  difficult task  and
does not substantially modify the  results reported here. 
Because of space limitations, the necessary modifications to 
account for slowly varying potentials will be presented 
elsewhere~\cite{unpub}. However, it is worth pointing out that in recent 
experiments~\cite{microchip} using 
microchip traps the confining potential could be tailored
to a shape very close to that of a square well  
with very high barriers. 
Thus we expect that  our results are directly relevant to future 
experiments in that type of traps,  as well as to the $^4$He systems 
considered in Refs.\cite{Wada01,Calbi01}

To study the low-energy behavior of a bosonic 1d 
{\it quantum} liquid, we rely on
Haldane's harmonic-fluid approach~\cite{Haldane81b}. 
How to generalize this approach to deal with boundaries
is explained below. In general, this method is able to account 
for the long wave-length properties of the system. In this particular 
case, this means that we cannot describe the  properties very 
close to the boundaries,  but for most experiments 
this is not  as important  as the effect of the boundaries 
on the bulk properties. With these provisos,  
the low-energy effective Hamiltonian 
as well as  asymptotic expressions for the ground 
state density and one-particle density matrix will be derived in this letter. 
We also analyze  finite-size and boundary effects  on 
the momentum distribution. This quantity is experimentally
accessible~\cite{Girardeau01,Mahan00}. 
Finally, we show that the 
quadratic character of the effective 
Hamiltonian allows us to derive 
Bijl-Jastrow  wave function
for the ground state~\cite{Pham9901}. Numerical 
evidence (e.g. Ref.~\cite{Affleck01})  suggests 
that our results should apply to 
systems containing even a few tens of particles.

 We begin by assuming that the system is confined by a very anisotropic potential
and that temperature is much lower than the excitation gap for
the transverse degrees of freedom 
(a more detailed analysis of the conditions for (quasi)
one-dimensionality can be found in
~\cite{Monien98,Yip01, Wada01, Calbi01,Olshanii98,Petrov00}).
$N_{\rm o}$ spinless bosons in the lowest transverse level of the
confining potential will be described by the following
Hamiltonian,
\begin{equation}\label{eq0}
H = \frac{\,\hbar^2}{2M} \int_{0}^{L} dx \: 
\partial_x \Psi^{\dagger}(x) \partial_x \Psi(x) 
 + \frac{1}{2}  \int_{0}^{L} dx \int_{0}^{L} 
dx^\prime \: v(x-x^\prime) \rho(x) \rho(x^\prime),
\end{equation}
where  $M$  is boson mass, 
$\rho(x) = \Psi^{\dagger}(x) \Psi(x)$ is the density operator, 
and  $[\Psi(x), \Psi^{\dagger}(x^\prime)] = \delta(x-x^\prime)$, 
but otherwise commute as corresponds
to bosons. The interaction $v(x)$ can be totally 
general as long as  it has a short-range repulsive 
part, and it does not  decay slower than $1/|x|^2$.
Confinement in the longitudinal direction, $x$,
is described by imposing that the field operator vanishes
at the boundaries, i.e. $\Psi(x) = 0$ for $x=0, \: L$, where $L$
is the system size. As discussed above, apart from this effect, 
the longitudinal confining potential  is neglected in what follows.

  For $v(x) = g \: \delta(x)$ the model was exactly diagonalized  by Lieb 
and Liniger~\cite{Lieb63a}. Lieb~\cite{Lieb63b} also showed
that the low-energy excitation spectrum  for $g>0$ is adiabatically 
connected with that of a system of 
impenetrable bosons (the Tonks gas~\cite{Tonks36}, 
where $g\to +\infty$). Girardeau~\cite{Girardeau60} 
had previously found that   the Tonks gas and 
a system of free spinless fermions have the same 
spectrum. The latter presents a 
Fermi ``surface'' consisting of two points, and 
the excitations are particle-hole
pairs. The collective modes that exhaust the 
low-energy spectrum correspond to coherent 
superpositions of particle-hole pairs.
Quite generally, a sufficiently short-ranged potential 
will make particles effectively impenetrable 
when their relative energy is small. Therefore, it seems
reasonable to expect that the description in terms of 
the collective modes remains good for a fairly large class of 
models. Yet, different models have
different high-energy structure and this  also leads to sizable corrections
to the sound velocity and long-distance correlations of the Tonks gas.

  To make the discussion more quantitative, 
we  follow Ref.~\cite{Haldane81b} 
and work in the density-phase representation 
of the boson field operator,
$\Psi^{\dagger}(x) = \sqrt{\rho(x)} \: e^{-i\phi(x)}$,
where $[\rho(x),\: e^{-i\phi(x')}] = \delta(x-x') 
\: e^{-i\phi(x)}$. At low energies the matrix elements
of $\partial_x\phi(x)$, as
 well as deviations of $\rho(x)$ from the mean
density $\rho_{\rm o}$ ($\equiv N_{\rm o}/L$),  are small. Thus, 
if the  long wave-length density fluctuations are
represented as $\rho_{\rm o}  + \Pi(x)$, one has 
 $[\Pi(x), \phi(x')] = i \delta(x-x')$,
i.e. the fields $\Pi(x)$ and $\phi(x)$ are canonically 
conjugate. The Hamiltonian
in Eq.~(\ref{eq0}) can be linearized in terms 
of them to obtain an effective low-energy Hamiltonian,
\begin{equation}\label{eq2}
H_{\rm eff} = \frac{\hbar v_s}{2}\: \int_{0}^{L} dx \: \left[ 
\frac{\pi}{K} \Pi^2(x) + \frac{K}{\pi} 
\left(\partial_x \phi(x) \right)^2 \right],
\end{equation}
where $v_s$ and $K$ should be regarded as 
phenomenological parameters. They can
be determined either from an exact 
solution~\cite{Haldane81b,Haldane81c} 
(when available~\cite{Lieb63a,Lieb63b}), numerically~\cite{Affleck01},
or from the experiment~\cite{Wada01}. Once these
parameters have been obtained, the above
Hamiltonian provides a complete description
of the low-lying excitations, independently
of the details of $v(x)$~\cite{Haldane81a}.

 Next, we introduce another field $\theta(x)$ related to $\Pi(x)$
by $\partial_x \theta(x)/\pi = \rho_{\rm o} + \Pi(x)$,
which implies that $\theta(x)$ increases by $\pi$ every time 
$x$ surpasses one particle. This means that
\begin{equation}\label{eq4}
\frac{1}{\pi} \big[ \theta(L) - \theta(0) \big] =  N 
\end{equation}
counts the total number of particles in the system. 
Identifying particle positions  with 
the points where  $\theta(x)$ changes by $\pi$ also 
allows to construct a representation of the full 
density operator~\cite{Haldane81b},
$\rho(x) =  \partial_x \theta(x) \:  \sum_{n=-\infty}^{+\infty} 
\delta(\theta(x) - n\pi)$, which with the help of 
Poisson's summation formula can be written as
\begin{equation}\label{eq5}
 \rho(x) = \left[\rho_{\rm o} + \Pi(x) \right] \: 
\sum_{m=-\infty}^{+\infty} e^{2im\theta(x)}. \label{eq5b}
\end{equation}
The field operator is then given
(up to an overall prefactor~\cite{Haldane81b}) by
\begin{equation}\label{eq6}
\Psi^{\dagger}(x) \sim  \left[\rho_{\rm o} + \Pi(x) \right]^{\frac{1}{2}} \: 
\sum_{m=-\infty}^{+\infty} e^{2im\theta(x)} \: e^{-i\phi(x)}.
\end{equation}
This expression yields an operator that commutes at different points,
as can be checked by using the mode expansions given below. 
An anti-commuting operator, which 
represents a Fermi field, can be constructed~\cite{Haldane81b} as 
$\Psi^{\dagger}_F(x) = \Psi^{\dagger}(x) e^{i\theta(x)}$.
 The above expressions, 
Eqs.~(\ref{eq2}) to (\ref{eq6}), provide 
us with the tools to compute
the spectrum and  correlation functions. However,
we have not yet touched upon
the issue of  boundary conditions. For open boundary conditions (OBC's)
we must demand that the field operator vanishes at the boundary,
which implies that
\begin{equation}\label{eq7}
\sum_{m=-\infty}^{+\infty} e^{2im\theta(0)} = 0.
\end{equation}
By construction, this is so if 
$\theta(0) = \theta_{\rm o} \neq n \pi$, $n$ being
an integer (the precise value of real number 
$\theta_{\rm o} \neq n \pi$ will not be important in what follows); The 
boundary condition is also obeyed for $x = L$
since  Eq.~(\ref{eq4}) relates $\theta(L)$  to $\theta(0)$.
The following mode expansions for $\theta(x)$
and $\phi(x)$ are then obtained  ($q = m \pi/L$, for $m = 1, 2, 3, \ldots$)
\begin{eqnarray}
\theta(x) &=& \theta_{\rm o} + \frac{\pi x }{L}N   + i \sum_{q > 
0} \Big( \frac{\pi K}{q L} \Big)^{\frac{1}{2}}\: e^{-\alpha q /2}  
\sin(qx) \big[b(q) - b^{\dagger}(q)\big], \label{eq8a} \\
\phi(x) &=& \phi_{\rm o}   + \sum_{q > 0}  \Big( \frac{\pi}{q L K} 
\Big)^{\frac{1}{2}}\: e^{-\alpha q /2}
\cos(qx)  \big[ b(q)+   b^{\dagger}(q)\big], \label{eq8b}
\end{eqnarray}
which diagonalize the Hamiltonian in  Eq.~(\ref{eq2}): 
\begin{equation}\label{eq9}
H_{\rm eff} = \sum_{q > 0} \hbar \omega(q) \: 
b^{\dagger}(q) b(q) + \frac{\hbar\pi v_s}{2LK} (N-N_{\rm o})^2,
\end{equation}
where $\omega(q \ll \rho_{\rm o}) = v_s \:
q > 0$,  and $[b(q), b^{\dagger}(q')] = \delta_{q,q'}$,
commuting otherwise.  That is, the low-energy excitations 
are linearly dispersing standing ``phonons''.
The  cutoff $\alpha \sim \rho^{-1}_{\rm o}$ 
in Eqs. (\ref{eq8a},\ref{eq8b})
makes explicit that these expansions are only 
meaningful as long as we restrict ourselves to 
a low-energy subspace where the phonon
wave-length is much longer than 
$\rho^{-1}_{\rm o}$. Besides the phonons, one
can create excitations that change the number of
particles. These are described by a pair of
operators $(N,\phi_{\rm o})$, which obey $[N, e^{-i \phi_{\rm o}}]  = e^{-i\phi_{\rm o}}$.
In contrast to PBC's, which were used in~\cite{Haldane81b},  
only one pair of these operators is needed
because, as mentioned
in the introduction, one cannot excite quantized persistent flows
in a longitudinally confined system. Finally, from 
Eq.~(\ref{eq9}) it follows that the compressibility 
$\kappa = \rho^{-2}_{\rm o}(d\rho_{\rm o}/d\mu)$ 
is proportional to $K$. The
Tonks gas has $K = 1$ (free spinless fermions), 
while for weakly interacting bosons $K \to +\infty$. 

  Using Eqs.~(\ref{eq5b},\ref{eq8a}) we 
have computed the ground
state density for separations from the 
boundaries larger than $\rho^{-1}_{\rm o}$. 
To  leading order in each harmonic 
of $2\pi \rho_{\rm o}$ we obtain 
\begin{equation}\label{eq10}
\langle \rho(x) \rangle  = \rho_{\rm o} 
\left\{ 1 +  \sum_{m=1}^{+\infty} 
A_{m} \: \frac{\cos(2m \pi  \rho_{\rm o}x + 
\delta_m)}{\left[\rho_{\rm o} d(2 x)\right]^{m^2 K} } \right\},
\end{equation}
where  $d(x) = (2L/\pi) \left| \sin(\pi x/2L) \right|$, and
 $A_m$ and $\delta_m$ are model dependent 
coefficients~\cite{coeffs}. 
Since in 1d the distinction 
between bosons and fermions is 
blurred by interactions~\cite{Haldane81b},
the ground state density exhibits
Friedel oscillations characteristic of Fermi 
systems~\cite{Fabrizio95,Wang9600,Affleck01}. 
They are induced by
the boundaries, which break translational symmetry,
and therefore would be absent
if we had assumed PBC's. For sufficiently
strong repulsive interactions ($K < 2$~\cite{Haldane81b,Fisher89,unpub}),
these oscillations can be pinned by a periodic
potential (e.g. an optical lattice) of wave length equal to  $\rho^{-1}_{\rm o}$.
The system would thus undergo a transition to
a Mott insulating regime~\cite{Bloch02}.

 The presence of the boundaries also  modifies   
the structure of  correlation functions. We have computed
the one-particle density matrix at zero temperature. 
Far from the boundaries
and for $|x-x^\prime| \gg \rho^{-1}_{\rm o}$, the result reads
\begin{eqnarray}
\langle \Psi^{\dagger}(x) \Psi(x') \rangle &\sim&  \rho_{\rm o} 
\left[ \frac{\rho^{-1}_{\rm o}
\sqrt{d(2x) d(2x')}}{d(x+x') d(x-x')}\right]^{\frac{1}{2K}}
 \sum_{m,m'=-\infty}^{+\infty} C_{m,m'}\: 
e^{-i (m+m')\pi {\rm sgn}(x-x')/2}\nonumber \\
&\times& \left[ \frac{d(x+x')}{d(x-x')}\right]^{2mm'K}   
 \frac{  e^{2\pi i \rho_{\rm o }  (m x - m' x')}  } 
{\left[\rho_{\rm o} d(2 x) \right]^{m^2 K} 
\left[\rho_{\rm o} d(2 x') \right]^{m'^2 K}}. \label{eq11}
\end{eqnarray}
Again $B_{m,m'}, C_{m,m'}$ are model
dependent (complex) coefficients which cannot be 
determined by this method. Bulk behavior  
is recovered for $ |x-x^\prime| \ll   \min\{x,x^\prime,L-x, L-x^\prime\}$. 
In this limit, the less oscillatory  terms in Eq.~(\ref{eq11}) 
are those where $m=m'$, which yield the 
expression in Ref.~\cite{Haldane81b}~\cite{notation}.
Using the same formalism other correlation functions,
such as the density correlation function,  have been obtained. These
will reported elsewhere~\cite{unpub} as only the density matrix, 
Eq.(\ref{eq11}), is needed here.

 From the one-particle density matrix 
the momentum distribution can be obtained~\cite{Eggert96}:
\begin{equation}\label{eq12}
n(p,L) =\frac{1}{L} \int_{0}^{L} dx \int_{0}^{L} dx' \: 
\langle \Psi^{\dagger}(x) \Psi(x')\rangle \: e^{ip(x-x')},
\end{equation}
which we have normalized so that 
$(2\pi  \rho_{\rm o})^{-1}\int dp \, n(p)   = 1$. The dominant contribution
at small momentum is given by the $m=m'=0$ term   in Eq.~(\ref{eq11}),
henceforth denoted $f(x,x',L)$.
In the thermodynamic limit $L \to \infty$,  this function
decays as $|x-x^{\prime}|^{-1/2K}$, which
implies that    $n(p \ll \rho_{\rm o}) \sim p^{-\beta}$ with 
$\beta = 1-(2K)^{-1}$.  However, when 
the system is finite these power laws do not hold. 
This poses  a problem to determine the exponent $\beta$
from experimental measurements  of the momentum 
distribution. Measuring the exponent 
is important, not only as a means to obtain $K$ 
and  get an idea of the strength of the interactions, 
but also because $\beta$ controls the decay of phase  correlations: 
For $\beta$ close to $1$ (i.e. $K \gg 1$) phase 
correlations will decay very slowly, and 
therefore the system will look like 
a Bose-Einstein condensate, even if, strictly speaking, 
in 1d bosons do not condense. On the other hand, when 
$\beta \approx 1/2$, the system will behave as a Tonks gas.
\begin{figure}[b]
\centerline{\resizebox{9cm}{!}{\includegraphics{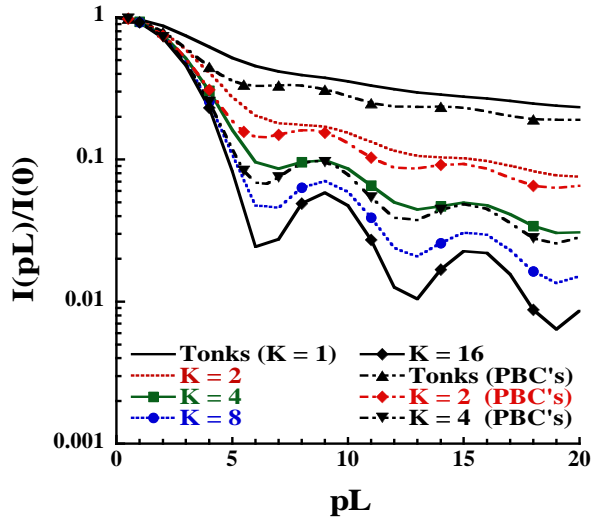}}}
\caption{Momentum distribution $n(p,L)$ vs. $pL$ ($L$ 
is the system's size), for different values
of $K$. It has been normalized to its value at 
$p = 0$ (Recall that  $n(0,L) = (\rho_{\rm o} 
L)^{\beta}\: I(0)$, with $\beta = 1 - (2K)^{-1}$. Hence  $n(p,L)/n(0,L) = 
I(pL)/I(0)$, cf. Eq.~(\ref{eq13})). Some results ($K=1,2,4$) with
periodic boundary conditions (PBC's)
are also shown for comparison.}
\label{fig1}
\end{figure}

 To find the exponent $\beta$, we must resort to a different type of  analysis.
Let us first note that the  function $f(x,x^\prime,L)$ is homogeneous, i.e. 
$f(x,x',L) = s^{1/2K} f(s x, s x', s L)$. Choosing $s = 1/L$, it follows that 
$f(x,x',L) = L^{-1/2K} F(x/L, x'/L)$, for 
some {\it scaling} function $F(\xi, \xi')$. Introducing this
result into Eq.~(\ref{eq12}), 
\begin{equation}\label{eq13}
n(p \ll \rho_{\rm o},L) = \left(\rho_{\rm o} L\right)^{\beta}\: I(pL),
\end{equation}
where $I(pL) = 2\rho^{-\beta}_{\rm o}\int_{0}^{1} d\xi \int_{0}^{1} d\xi' \: 
F(\xi, \xi') \cos[pL(\xi-\xi')]$. This function has 
been plotted in Fig. \ref{fig1} for different values of $K$.
Eq.~(\ref{eq13}) thus implies that the exponent $\beta$ can be
extracted from a finite-size scaling analysis of the experimental data for
$n(0,L)$ collected in systems of different size but equal density, $\rho_{\rm o}$. 
Girardeau and Wright~\cite{Girardeau01} 
have proposed a  non-destructive 
method to measure $n(p,L)$. With this technique, one can try to  
obtain $n(p \to 0, L)$ as a function of $L$, using a blue-detuned
laser beam to split a quasi 1d trapped atomic gas 
into pieces of smaller size which could be probed independently. 
Real systems, however, 
are at finite temperature, whereas the previous analysis is for  $T=0$. 
At finite temperature, correlations decay 
exponentially with distance~\cite{Haldane81a}.
Repeating the above analysis $n(p, L, T) = 
(\rho_{\rm o} L)^{\beta} I'(pL,TL/\hbar v_s)$. Thus, when
measuring systems of different size the product $TL$ must be kept constant.
However, for $T < \hbar v_s/L_{\rm max}$, 
where $L_{\rm max}$ is the size of the largest
system considered, quantum fluctuations 
dominate over thermal fluctuations
and the dependence on $T$ can be neglected so that our $T=0$ analysis 
should be valid.

  Finally, we derive a Bijl-Jastrow wave function
for the ground state of $N_{\rm o}$ interacting 
bosons confined in a box with open boundaries.
For the Tonks gas, it was pointed out by 
Girardeau~\cite{Girardeau60} that the exact ground
state is of Bijl-Jastrow form.
Recently, Pham {\it et al.}~\cite{Pham9901} 
have remarked that the quadratic form of the effective
Hamiltonian, Eq.~(\ref{eq2}), implies that a 
Bijl-Jastrow wave function is a good
approximation to the ground state. Here we extend 
their derivation to OBC's.
Expanding $\Pi(x) = \Pi_{\rm o}/L + 
\sqrt{2/L} \sum_{q>0} \hat{\Pi}_{q} \cos(qx)$
and $\phi(x) = \phi_{\rm o} + \sqrt{2/L} \sum_{q>0} \hat{\phi}_{q} \cos(qx)$, 
with $[\hat{\Pi}_q,\hat{\phi}_{q'}] = i \delta_{q,q'}$ 
and $\Pi_{\rm o} = N-N_{\rm o}$, 
$H_{\rm eff}$ is given by
\begin{equation}\label{eq14}
H_{\rm eff} =  \frac{\hbar \pi v_s}{2K} \sum_{q > 0} \left[\Pi^2_q 
 - \left(\frac{qK}{\pi}\right)^2 \frac{\delta^2}{\delta\Pi^2_q}\right] +
\frac{\hbar \pi v_s}{2LK} \Pi_{\rm o}^2, 
\end{equation}
where   $\hat{\phi}_q$ has been replaced by  $-i \delta/\delta\Pi_q$, as 
implied by its commutator with $\hat{\Pi}_q$.
Thus the ground state  is a gaussian 
in momentum space,  $\Phi_{\rm o} = \exp \left\{-\pi 
\sum_{q>0}{\Pi^2_q/2qK}\right\}$,  which in real space reads
\begin{eqnarray}
\Phi_{\rm o}(x_1,\ldots, x_N) &=& \exp \left\{\frac{1}{2} \int_{0}^{L}
 dx \int_{0}^{L} dx' \: \rho(x)  {\cal K}(x,x') \rho(x') \right\} 
\nonumber \\ 
&\propto&  \prod_{i=1}^{N_{\rm o}} 
\left[\frac{d(2x_i)}{\pi^{-1}L}\right]^{\frac{1}{2K}} 
 \prod_{i<j} \left[ \frac{d(x_i+x_j) 
d(x_i-x_j)}{\pi^{-2} L^2} \right]^{\frac{1}{K}},
 \end{eqnarray}
where we have replaced $\Pi(x)$ 
by $\rho(x) - \rho_{\rm o}$, and then  
used  $\rho(x) = \sum_{i=1}^{N_{\rm o}} \delta(x-x_i)$ 
and  ${\cal K}(x,x') =  \log\left|L^{-2} \pi^2 d(x+x') d(x-x') \right|/K$.
The result bears some resemblance to the  exact ground states of 
the Tonks gas~\cite{Girardeau01b} ($K = 1$) and the 
Calogero-Sutherland~\cite{Sutherland98} models 
in a harmonic trap.  The differences are due to the different
confining potential considered.  As in those cases, however,
the wave function vanishes when two particles approach each
other, revealing strong correlations between the bosons.

  In conclusion, we have extended 
the harmonic-fluid approach~\cite{Haldane81b} to obtain 
the low-lying  excitations and study finite-size and boundary effects 
in  systems of  interacting bosons in 1d that are
confined longitudinally.  When analyzing their properties, 
these effects must be taken  into account. 
This has been illustrated by considering the momentum 
distribution at small momentum, which does not behave as the power law predicted for
the infinite system. Finally, we have also derived
a Bijl-Jastrow  wave function for the ground state.
A more complete   account will be given elsewhere~\cite{unpub}.


  It is a pleasure to thank Pedro M. Echenique 
for his kind hospitality at the DIPC in Donostia.
I am also grateful  to Michele Fabrizio, Angel Garc\'\i a-Adeva,  
Yu Lu, Fernando Sols,  Erio Tosatti, and
especially to Andrew Ho and Alexander Nersesyan, 
for  useful conversations and comments.


\begin{thebibliography}{30}
%
\bibitem{Haldane81a}
F.~D.~M. Haldane, J.~Phys.~C {\bf 14}, 2585 (1981).
%
\bibitem{Haldane81b}
 F.~D.~M. Haldane, Phys.~Rev.~Lett.~{\bf 47}, 1840 (1981).
%
\bibitem{Gorlitz01}
A. G\"orlitz {\it et al.}, Phys.~Rev.~Lett.~{\bf 87}, 130402 (2001).
%
\bibitem{Wada01}
N. Wada {\it et al.}, Phys.~Rev.~Lett.~{\bf 86}, 4322 (2001).
%
\bibitem{Calbi01}
M.~M. Calbi {\it et al.}, Rev.~ Mod.~ Phys.~{\bf 73}, 857 (2001).
%
\bibitem{microchip}
W. H\"ansel, P. Hommelhoff, T. W. H\"ansch, and J. Reichel, Nature,
{\bf 413}, 498 (2001).
%
\bibitem{Monien98}
H. Monien, M. Linn, and N. Elstner, Phys.~Rev.~A {\bf 58}, R3395 (1998).
%
\bibitem{Yip01}
S.~K. Yip, Phys.~Rev.~Lett.~{\bf 87} 130401 (2001).
%
\bibitem{boundaries}
In the system   of  Ref.~\cite{Wada01} the pores are
$ 3000$ \AA~ long. This should be 
compared with the density,
$\rho_{\rm o} \approx 1$ \AA$^{-1}$. 
Estimates of  the max. length attainable in
traps of the type used in Ref.~\cite{Gorlitz01} give
$\approx 10^7$ \AA~ with  
$\rho_{\rm o} \sim 10^{-4}$ \AA$^{-1}$~\cite{Monien98}. 
%
\bibitem{Fabrizio95}
M.~ Fabrizio and A.~O. Gogolin, Phys.~Rev.~ B {\bf 51}, 17827 (1995).
%
\bibitem{Eggert96}
S. Eggert, H. Johannesson, and A. Mattsson, 
Phys. Rev. Lett. {\bf 76}, 1505 (1996).
%
\bibitem{Wang9600}
Y. Wang, J. Voit, and F.-C. Pu, Phys.~Rev.~B {\bf 54}, 8491 (1996).
%
\bibitem{Affleck01}
S. Eggert and I. Affleck, Phys.~Rev.~B {\bf 46}, 10866 (1992);
S.~R. White, I. Affleck, and D.~J. Scalapino, Phys.~Rev.~B, {\bf 65},
165122 (2002).
%
\bibitem{Olshanii98}
M.~ Olshanii, Phys.~Rev.~Lett.~{\bf 81}, 939 (1998).
%
\bibitem{Petrov00}
D.~S. Petrov {\it et al.}, Phys.~Rev.~Lett. {\bf 85}, 3745 (2000).
%
\bibitem{Girardeau01}
M.~D. Girardeau and E.~M. Wright, Phys.~Rev.~Lett.~{\bf 87}, 050403 (2001).
%
\bibitem{Mahan00}
G.~D. Mahan, {\it Many-Particle Physics}, Plenum (New York, 2000), pag. 697. 
%
\bibitem{Pham9901}
K.~V. Pham, M. Gabay, and P. Lederer, 
Eur.~ Phys.~J.~ B {\bf 9}, 573 (1999); Phys.~Rev.~B~{\bf 61}, 16397(2001).
%
\bibitem{Lieb63a}
E.~H. Lieb and W.~Liniger, Phys.~Rev.~{\bf 130}, 1605 (1963).
%
\bibitem{Lieb63b}
E.~H. Lieb, Phys.~Rev.~{\bf 130}, 1616 (1963).
%
\bibitem{Tonks36}
L.  Tonks, Phys.~Rev.~{\bf 50}, 955 (1936).
%
\bibitem{Girardeau60}
M. Girardeau, J.~Math. Phys.~ (N.Y.) {\bf 1}, 516 (1960).
%
\bibitem{Haldane81c}
F.~D.~M. Haldane, Phys.~Lett.~{\bf 81}A, 153 (1981).
%
\bibitem{coeffs}
E.g. for the Tonks gas ($K=1$) confined by hard walls, 
one obtains $\delta_1 = \pi/2$, $A_1 = 1/\pi$, and $A_{m} = 0$ for $m > 1$.
%
\bibitem{Fisher89}
M.~P.~A. Fisher {\it et al.},
Phys.~Rev.~B, {\bf 40}, 546 (1989).
%
\bibitem{Bloch02}
M. Greiner {\it et al.}, Nature, {\bf 415}, 39 (2002).
%
\bibitem{notation}
Haldane denotes $\eta \equiv 2 K$.
%
\bibitem{Girardeau01b}
M.~D. Girardeau, E. M. Wright, and J. M. Triscari,
Phys.~Rev.~ A {\bf 63}, 033601 (2001).
%
\bibitem{Sutherland98}
B. Sutherland, Phys.~Rev.~Lett.~{\bf 80}, 3678 (1998).
%
\bibitem{unpub}
M.~A. Cazalilla, in preparation.
%
\end{thebibliography}
\end{document}